\documentclass[12pt]{article}
\usepackage{amsfonts}
\usepackage{amssymb}
\usepackage{amsmath}

\input epsf
\input{tcilatex}
\begin{document}

\thispagestyle{empty}

\begin{center}
\null\vspace{-1cm} \hfill \\[0pt]
\vspace{1cm}\medskip {\large \textbf{F-Susy And The Three States
Potts Model}}

\vspace{1.5cm} \textbf{M.B. Sedra} , \textbf{J. Zerouaoui} \\[0pt]
{Universit\'{e} Ibn Tofail, Facult\'{e} des Sciences,
D\'{e}partement de
Physique,\\[0pt]
Laboratoire de Physique de la Mati\`{e}re et Rayonnement (LPMR), K\'{e}%
nitra, Morocco.}\\[0pt]
\end{center}

\begin{abstract}
In view of its several involvements in various physical and
mathematical contexts, 2D-fractional supersymmetry (F-susy) is
once again considered in this work. We are, for instance,
interested to study the three states Potts model $(k = 3)$ which
represents with the tricritical Ising model $(k = 2)$ the two
leading examples of more general spin $1/k$ fractional
supersymmetric theories.
\end{abstract}

\newpage

\section{Introduction}

Fractional spin symmetries (FSS) \cite{ref 1, ref 2, ref 3}, which
deals with exotic particles, is an important subject that emerges
remarkably twenty years ago in coincidence with the growing
interest in high energy and condensed matter physics through
quantum field theory \cite{ref 4}, conformal symmetries \cite{ref
5} and string theory \cite{ref 6}. \\\\
These symmetries are specific for two-dimensional theories and
play a pioneering role in the study of $D=2$ conformal field
theories and integrable $\Phi _{1,3}$ deformation of minimal
models \cite{ref 7}. Well known examples are given by the standard
$D=2$ supersymmetry generated by spin $1/2$ charge operators
$Q_{\pm 1/2}$ and the superconformal symmetry exhibiting an
infinite number of half integer constants of motion \cite{ref
8}.\\\\
Other non common examples are the $c=1-\frac{6}{p(p+1)}$ minimal
models containing among their $\frac{p(p-1)}{2}$ primary fields a
spin $\frac{(p+2)}{2}$ conformal field which, combined with the
energy momentum tensor, generate a kind of generalized
superconformal symmetry. \\\\Recall that the usual superconformal
invariance is generated by a spin $3/2$ conserved current in
addition to the Virasoro current of spin $2$ which give rise, in
Laurent modes, to the Neveu-Schwarz and Ramond superconformal
algebras. On the other hand, following \cite{ref 9}, the
$\phi_{1,3}$ deformation of the $c=6/7$ tricritical three states
Potts model exhibits a similar behavior as the $\phi_{1,3}$
deformation of the $c=7/10$ tricritical Ising model \cite{ref 10}.
Both of them admit fractional spin constants of motion namely
$Q_{\pm 1/3}^{\pm}$ and $Q_{\pm 1/2}$ surviving after the
perturbation and satisfying
\begin{equation}
Q_{\pm 1/2}^{2}=P_{\pm 1},
\end{equation}
and
\begin{eqnarray}
Q_{1/3}^{-3}&=&Q_{1/3}^{+3}=P_1\\
Q_{-1/3}^{-3}&=&Q_{-1/3}^{+3}=P_{-1}
\end{eqnarray}
where $P_{\mu}=(P_1, P_{-1})$ is the two dimensional energy
momentum vector.  The $\pm 1/3$ lower indices and $\pm$ upper ones
carried by the $Q_{\pm 1/3}$'s are respectively the values of the
spin and the charges of the $Z_3$ automorphism symmetry of
eqs(2-3). Note also that the above equations are particular
examples of more general fractional spin $s=\pm 1/k $ equations
generalizing the $D=2$ supersymmetry algebra and reading as
\begin{equation}
Q_{s}^{\pm k}=P_{ks}
\end{equation}
Setting $k=2$ and $k=3$, one gets respectively the standard $D=2
N=1$ supersymmetry and the leading generalized one eqs(2-3).
Moreover, it is established that under the $\phi_{1, 3}$
deformation, a $c(p)$ minimal model flows to the subsequent
$c(p-1)$ conformal one \cite{ref 7} in agreement with the
Zamolodchikov c-theory \cite{ref 11} according to which the
central charge $c$ is a decreasing function in the space of
coupling parameters. \\\\We focus in this work to renew our
interest in fractional supersymmetry, although several productions
have been made previously, since we believe that much more
important results can be extracted. This is also important to give
new breath to supersymmetry, conformal invariance and integrable
models as one of the best issues in the history of theoretical and
mathematical physics.

We present the $D=2$ three states Potts model as been the first
non trivial physical model corresponding to a fractional
deformation of the standard supersymmetry and show later how the
$D=2 (1/3 , 1/3)$ can be it's underlying invariance. We are
projecting through this first presentation among a series of
fourth coming works to shed new insights towards understanding
well these exotic symmetries and their possible incorporation in
various modern topics of theoretical and mathematical physics.

\section{$D=2$ Three states Potts Model}

It's now well known that the $c=6/7$ conformal theory and more
particularly the three critical Potts model (TPM), admits several
several infinite dimensional symmetries. The first kind is given
by the conformal symmetry whose generator $T_{\mu \nu}, \mu,
\nu=z, \bar z$ is nothing but the spin $2$ energy momentum tensor
which is symmetric and traceless.
The second infinite symmetry is generated by the so called conserved $W$%
-currents. Combined with the spin $2$-conformal current, the
$W$-currents generate a huge infinite symmetry known as the
$W$-symmetry \cite{ref 12}.\\\\ The famous example of this non
standard symmetry is given by the Zamolodchikov algebra generated
by $T$ of spin 2 and $W$ of spin 3 conformal currents. The common
property of $W$-symmetries is the fact that they are associated to
a non standard Lie algebra structure. $W$-algebra is a non linear
symmetry containing the conformal one as a particular case and is
known as the mediator of the integrability of the $\phi_{1,2}$
magnetic deformation of the $c=6/7$ critical theory \cite{ref
9}.\\\\
The third class of infinite symmetries of the TPM is generated by
fractional spin $4/3$ conserved currents $G_{4/3}^{\pm}$ and
${\bar G}_{4/3}^{\pm}$ in addition to the energy momentum tensor.
We shall refer to this symmetry as the $4/3$-superconformal
symmetry in analogy with the standard superconformal symmetry
generated by spin $3/2$ current and which we denote as spin
$3/2$-superconformal symmetry. \\\\This invariance of the $c=6/7$
critical theory generalizes in some sense the $N=1$ spin $3/2$
superconformal symmetry of the tricritical Ising model (TIM)
having central charge $c$ equal to $7/10$. Recall that TIM and TPM
are respectively given by the fourth and sixth levels of the
minimal series
\begin{equation}
c(p)=1-\frac{6}{p(p+1)}; p=3, 4,...,
\end{equation}
They appear also as the leading conformal theories of the $N=1$ spin $%
3/2$ superconformal discrete series
\begin{equation}
c(m)=\frac{3}{2}( 1-\frac{8}{m(m+2)}); m=3, 4,...,
\end{equation}
and the spin $4/3$ superconformal discrete one
\begin{equation}
c(m)=2( 1-\frac{12}{m(m+4)}); m=3, 4,...,
\end{equation}
Note also that eqs(5-7) may be regrouped into a two integers
discrete series as \cite{ref 13}
\begin{equation}
c(m, n)=\frac{3n}{n+2}( 1-\frac{2(n+2)}{m(m+n)}); n=1, 2,...,
\end{equation}
For $n=1$, one recover the unitary minimal models see eq(5). Putting $n=2$ and $%
n=4 $ in the above relation by keeping the integer $m$ free, we obtain
respectively the spin $3/2$-superconformal and $4/3$-superconformal
theories. However, letting the integer $n$ free and taking $m=3$ we get
\begin{equation}
c(3, n)=1-\frac{6}{(n+2)(n+3)}; n=1, 2,...,
\end{equation}
which corresponds to the minimal unitary series eq(5) once we set
$n+2=p$.\\
From this overlapping of the above discrete series, we deduce that
$N=0$ conformal models of eq(5) admit extra symmetries since they
appear as special critical models of the $c(m,n)$ theories.\\\\
Another interesting aspect exhibited by the TIM and TPM and more generally the $c(k)=1-\frac{3}{k(2k+1)}%
;k=2,3,...,$ conformal models is the integrability of their $\phi
_{1,3}$ deformation, see the last ref. in \cite{ref 7}. It's shown there that the thermal perturbation of the $%
c=6/7$ model induces an off critical spin $1/3$ supersymmetric
algebra surviving after the $\phi _{1,3}$ perturbation. This is a
finite dimensional symmetry generated by conserved charges
$Q_{+1/3}^{\pm }$ and $Q_{-1/3}^{\pm }$ carrying fractional spin
$s=\pm 1/3$ and non vanishing $Z_{3}$ charges, we have
\begin{eqnarray}
Q_{1/3}^{-3} &=&P_{1},Q_{-1/3}^{-3}=P_{-1} \\
Q_{1/3}^{+3} &=&P_{1},Q_{-1/3}^{+3}=P_{-1}
\end{eqnarray}
and
\begin{eqnarray}
Q_{1/3}^{-}Q_{-1/3}^{-}-{\bar{q}}Q_{-1/3}^{-}Q_{1/3}^{-} &=&\Delta
^{(-,-)}
\\
Q_{1/3}^{+}Q_{-1/3}^{+}-{q}Q_{-1/3}^{+}Q_{1/3}^{+} &=&\Delta ^{(+,+)} \\
Q_{1/3}^{-}Q_{-1/3}^{+}-{\bar{q}}Q_{-1/3}^{+}Q_{1/3}^{-} &=&\Delta ^{(-,+)}
\\
Q_{1/3}^{+}Q_{-1/3}^{-}-{q}Q_{-1/3}^{-}Q_{1/3}^{+} &=&\Delta ^{(+,-)}
\end{eqnarray}%
where $P_{3}$ is the usual $2d$ energy momentum vector and $\Delta
^{(r_{1},r_{2})},r_{1},r_{2}=\pm 1$ are topological charges. the parameter $%
q $ is such that $q^{3}=1$ chosen as $q=exp(2i\pi /3)$. It can be
thought of as the deformation parameter of the $U_{q}(sl(2))$
quantum enveloping algebra of $sl(2)$ \cite{ref 14}. The parameter
$q$ describes also the generator of the $Z_{3}$ discrete abelian
group. Denoting by $S^{\star }$ the critical action of TPM and by
$S$ its deformation $\phi _{h,\bar{h}}=\phi _{1,3}\otimes \phi
_{1,3};h,\bar{h}=5/7$ namely
\begin{equation}
S=S^{\ast }+\lambda \int d^{2}z \phi _{5/7, 5/7}
\end{equation}
where $\lambda $ is the perturbation parameter, it was shown that
the algebra eq(10-15) is a symmetry of the above deformed theory.
The conserved charges $Q_{s}^{\pm}$, $\Delta^{(r_1, r_2)}$ and
$P_{3s}, s=\pm 1/3 $ are realized as follows
\begin{eqnarray}
Q_{1/3}^{\pm } &=&\int \left[ dzG^{\pm }(z,{\bar{z}})+d{\bar{z}}\Gamma ^{\pm
}(z,{\bar{z}})\right] \\
Q_{-1/3}^{\pm } &=&\int \left[ d{\bar{z}}{\bar{G}}^{\pm }(z,{\bar{z}})+dz{%
\bar{\Gamma}}^{\pm }(z,{\bar{z}})\right]
\end{eqnarray}
and
\begin{eqnarray}
 \Delta ^{(+,+)} &=&\int
\left[ dz\partial ++d{\bar{z}}\bar{\partial}\right]
\varphi ^{(+,+)} \\
\Delta ^{(-,-)} &=&\int \left[ dz\partial +d{\bar{z}}\bar{\partial}\right]
\varphi ^{(-,-)} \\
\Delta ^{(+,-)} &=&\int \left[ dz\partial +d{\bar{z}}\bar{\partial}\right]
\varphi ^{(+,-)} \\
\Delta ^{(-,+)} &=&\int \left[ dz\partial +d{\bar{z}}\bar{\partial}\right]
\varphi ^{(-,+)} \\
\end{eqnarray}
and\begin{eqnarray}
P_{_{1}} &=&\int \left[ dzT+d{\bar{z}}\Theta \right] \\
P_{_{-1}} &=&\int \left[ d{\bar{z}}\bar{T}+dz\Theta \right]
\end{eqnarray}
The conformal field $\phi_{h, \bar h}^{(r, \bar r)}(z, \bar
z)=\phi_{h}^{r}(z)\otimes \phi_{\bar h}^{(\bar r)}(\bar z)$, with
$r, \bar r= 0,1,2$$(mod 3)$ are the $Z_3 \times \bar Z{_3}$ (left
right) charges, appearing in the above equations, which are built
as: the $4/3$ supersymmetric currents $G^\pm_{4/3}$ (resp ${\bar
G}^\pm_{-4/3}$) which carry only a left (resp. right) $Z_3$ charge
read as
\begin{equation}
G=\phi _{4/3,0}^{(\pm ,0)} , \bar G^{\pm}=\phi _{0, 4/3}^{(0,
\pm)}
\end{equation}
The magnetic order parameter fields $\varphi^{(+, +)}$ and its conjugate $%
\varphi^{(-, -)}$ are given by
\begin{equation}
\varphi^{(+, +)}=\phi _{1/21,1/21}^{(+, +)} , \varphi^{(-,
-)}=\phi _{1/21, 1/21}^{(-, -)}
\end{equation}
They carry the same left and right $Z_3$ charge contrary to the
magnetic disorder parameter fields $\varphi^{(+,-)}$ and
$\varphi^{(-,+)}$ which read as
\begin{equation}
\phi^{(+, -)}=\phi _{1/21,1/21}^{(+, -)} , \phi^{(-, +)}=\phi _{1/21,
1/21}^{(-, +)}.
\end{equation}
The remaining relevant fields of the TPM involved in eqs(17-25)
are
\begin{equation}
{\bar \Gamma}^{\pm}_{-2/3}=\phi _{1/21,1/21}^{(\pm, 0)} ,
\Gamma^{\pm}_{2/3}=\phi _{5/7,1/21}^{(0, \pm)} , D^0=\phi _{5/7,5/7}^{(0, 0)}
\end{equation}
The field $\Theta$ appearing in the two last relations eqs(24-25)
is the trace of the conserved energy momentum tensor of the off
critical theory. It measures the violation of the scale invariance
of the $\phi _{5/7,5/7}$ deformation of the TPM model. It reads
then as
\begin{equation}
\Theta\sim\lambda \phi _{5/7,5/7}^{(0, 0)}
\end{equation}
Note that the fields $\phi, \Gamma, \bar \Gamma$ and $D$
eqs(27-29) have values
of the spin $s=h-\bar h$ respectively equal to $0, (1-s), -(1-s), 0$ with $%
s=1/3$. Note also that the above field operators share some basic features
with the four fields involved in the $N=1$ spin $1/2$ supersymmetric $%
\phi_{1,3}=\phi_{3/5, 3/5}$ deformation of the TIM \cite{ref
10}.\\
There, these conformal fields have respectively the spin values $%
0,(1-s),-(1-s)$ and $0$ with $s=1/2$. They belong to the scalar
representation of the two dimensional $N=1$ spin $1/2$ supersymmetric
algebra
\begin{eqnarray}
Q_{s}Q_{s}+Q_{s}Q_{s} &=&P_{2s},s=\pm 1/2 \\
Q_{s}Q_{-s}+Q_{-s}Q_{s} &=&\Delta
\end{eqnarray}%
This algebra is generated by hermitian charges and admits a field
representation analogous to the field realization eqs(24-25) of
the off critical spin $1/3$ superalgebra. We have
\begin{eqnarray}
Q_{1/2} &=&\int dzG+d{\bar{z}}\Gamma ] \\
Q_{-1/2} &=&\int \left[ d{\bar{z}}{\bar{G}}+dz{\bar{\Gamma}}\right] \\
\Delta &=&\int \left[ dz\partial ++d{\bar{z}}\bar{\partial}\right] \varphi \\
P &=&\int \left[ d{z}T+d{\bar{z}}\Theta \right] \\
\bar P &=&\int \left[ d{\bar{z}}\bar T+dz\Theta \right]
\end{eqnarray}
We can define the following fields operators $G_{3/2}, {\bar G}%
_{-3/2}, \phi, F, G_{1/2}$ and $\bar G_{-1/2}$ in terms of the
field $\phi$ as follows
\begin{eqnarray}
G_{3/2} &=&\phi _{3/2,0},\bar{G}_{-3/2}=\phi _{0,3/2} \\
\varphi &=&\phi _{1/10,1/10},F=\phi _{3/5,3/5} \\
\bar{G}_{-1/2} &=&\phi _{1/10,3/5},G_{1/2}=\phi _{3/5,1/10}
\end{eqnarray}%
and where $\Theta $ is  shown also to be proportional to the
perturbation field $F= \phi_{3/5, 3/5}$ namely $\Theta =\alpha F$.
\section{The Underlying $D=2 (1/3, 1/3)$ Supersymmetry}
An important question that emerges when studying exotic fractional
symmetries is their superspace representations. In previous works
\cite{ref 2}, we succeeded to build representations of the $D=2$
fractional supersymmetric algebra, noted simply as $(1/k, 1/k)$
for $k=2,3,...$. The last notation indicates simply the left and
right-hand sides of the supersymmetric algebra generated by spin
$s=\pm 1/k$ charge operators $Q$ and $\bar Q$ satisfying
eq(4).\\\\
Based on this knowledge and on the fact that the particular choice
$k=2$ reproduce the standard $D=2$ supersymmetry, we focus in what
follows to show how $(1/3, 1/3)$ supersymmetric algebra can be
considered as the underlying symmetry of the $D=2$ three state
Potts model.\\\\
Recall first that there are few known models that exhibit the
$(1/3, 1/3)$ supersymmetric algebra. We quote the $C=6/7$ minimal
models and its $\phi_{1, 3}$ deformation. In these cases, the
$D=2(1/3, 1/3)$ superfields are characterized by their spins $s=
h-\tilde h$ and their scale dimension $\Delta=h+\tilde {h}$. They
contain $3\times 3$ component fields depending on the space
coordinates $z$ and $ \bar z$ and on the extra variable $u$
realizing the topological charge. Setting $h=\tilde {h}=1/21$, by
virtue of the $C=6/7$ minimal model \cite{ref 7}, the scalar
superfield $\phi_{1/21, 1/21}$ expands in $\theta _{-1/3}$ and
$\tilde {\theta} _{1/3}$ series as
\begin{equation}
\begin{array}{lcl}
 \phi_{({1/21, 1/21})}
 &=& \varphi_{({1/21,1/21})}+\theta_{-1/3}\psi_{({8/21,1/21})}+\tilde\theta_{1/3}\tilde\psi_{({1/21,
 8/21})}\\\\
 &+& \theta^{2}_{-1/3}\chi_{({15/21, 1/21})}+ \tilde\theta^{2}_{1/3}\tilde\chi^{2}_{({1/21, 15/21})}+
 \theta_{-1/3}\tilde\theta_{1/3}\xi_{({8/21, 8/21})}\\\\
&+&\theta^{2}_{-1/3}\tilde \theta
_{1/3}\lambda_{({15/21,8/21})}+\tilde\theta^{2}_{1/3}\theta
_{-1/3}\tilde\lambda_{({8/21, 15/21})}
+\theta^{2}_{-1/3}\tilde\theta ^{2}_{1/3}F_{({15/21, 15/21})}.
\end{array}
\end{equation}
From this expansion we recognize the fields involved in the TPM
discussed in the previous section. Note for instance the last term
of this expression namely $F_{({15/21, 15/21})}$ is nothing but
the $\phi_{1,3}$ field $\phi_{({5/7, 5/7})}$ of the TPM or $C=6/7$
conformal theory. This is a $D=2(1/3, 1/3)$ supersymmetric
invariant quantity exactly as for the field of the $D=2(1/2, 1/2)$
supersymmetric $C=7/10$ minimal model.
\\\\
Based on this presentation, others important aspects of fractional
supersymmetries will be presented in our forthcoming works.

\end{document}